\documentclass[aps,preprint]{revtex4}
\usepackage{amssymb}
\usepackage{amsmath}

\input{tcilatex}
\begin{document}

\title{Isolating the vortex core Majorana state in p-wave superconductors. }
\author{B. Rosenstein$^{1,2}$, I. Shapiro$^{3}$, B. Ya. Shapiro$^{3}$}
\affiliation{$^{1}$Department of Electrophysics, National Chiao Tung University, Hsinchu,
Taiwan, R.O.C. }
\affiliation{$^{2}$Applied Physics Department, Ariel University Center of Samaria, Ariel
40700, Israel\\
$^{3}$Department of Physics, Institute of Superconductivity, Bar-Ilan
University, Ramat-Gan 52900, Israel}

\begin{abstract}
The spectrum of core excitations of the Abrikosov vortex pinned by a
nanohole of the size of the coherence length is considered. While the
neutral zero energy Majorana core state remains intact due to its
topological origin, the energy of charged excitations is enhanced
significantly compared to that in the unpinned vortex. As a consequence of
the pinning the minigap separating the Majorana state\ from the charged
levels increases from $\Delta ^{2}/E_{F}$ for ($E_{F}$ is Fermi energy, $%
\Delta $ - the bulk $p$-wave superconducting gap) to a signicant fraction of 
$\Delta $. Suppression of the thermodynamic and kinetic effects of the
charged excitations allows to isolate the Majorana state so it can be used
for quantum computation. It is proposed that thermal conductivity along the
vortex cores is a sensitive method to demonstrate the minigap. We calculate,
using Butticker - Landauer - Kopnin formula, the thermal conductance beyond
linear response as function of the hole radius.
\end{abstract}

\maketitle

Spin-triplet $p$-wave superfluids, both neutral, such as liquid $He^{3}$
(and recently generated by the Feshbach resonance on $Li^{6}$ and $K^{40}$)
and charged such as superconducting material $Sr_{2}RuO_{4}$ and possibly
heavy fermion $UPt_{3}$ have resulted in very rich physics \cite{Leggett}.
The condensate is described by a generally tensorial complex order parameter 
$\Delta $ exhibiting great variety of the broken symmetries ground states.
The broken symmetry and boundary conditions give rise to the continuous
configuration of the order parameter as nontrivial topological excitations 
\cite{Volovik}. Especially interesting is the case of the so-called
topological superconductors, characterized by presence of electron-hole
symmetry and absence of both the time-reversal and spin-rotation symmetry.
Realizations of topological $p$-wave superfluids are chiral superconductors
like $Sr_{2}RuO_{4}$, with order parameter of the $p_{x}\pm ip_{y}$ symmetry
type \cite{Maeno} and the ABM - phase\cite{Leggett} of superfluid $He^{3}$
and other fermionic cold atoms \cite{BEC} and topological superconductor $%
Cu_{x}Bi_{2}Se_{3}$ that produces an equivalent pseudospin system on its
surface\cite{Ong}.

Magnetic field in type II superconductors easily creates stable line - like
topological defects, Abrikosov vortices\cite{Kopnin}, see Fig.1a. In the
simplest vortex the phase of the order parameter rotates by $2\pi $ around
the vortex and each vortex carries a unit of magnetic flux $\Phi _{0}$.
Quasiparticles near the vortex core "feel" the phase wind by creating a set
of discrete low-energy Andreev bound state. For the $s$-wave superconductors
when the vortices are unpinned (freely moving) these states were
comprehensively studied theoretically including the excitations spectrum 
\cite{Caroli}, density of states \cite{Gygi}, their role in vortex viscosity%
\cite{Kopnin}, contribution to the heat transport\cite{Kopnin03} and to the
microwave absorption \cite{Janko}. The low lying spectrum of quasiparticle
and hole excitations is equidistant, $E_{l}=l\omega ,$ where angular
momentum $l$ takes half integer values. The "minigap" in the $s$-wave
superfluids is of order of $\omega =\Delta ^{2}/E_{F}$ $<<\Delta $, where $%
\Delta $ is the energy gap and $E_{F}$ is the Fermi energy.

Free vortices in the $p$-wave superconductors exhibit a remarkable
topological feature of appearance of the zero energy mode in the vortex core 
\cite{Volovik99}. The spectrum of the low energy excitations remains
equidistant, $E_{l}=\left( l-1\right) \omega $, but now $l$ is integer\cite%
{Sigrist}. The zero mode represents a condensed matter analog of the
Majorana fermion first noticed in elementary particle physics\cite{Wilczek}.
Its remarkable feature is linked to the fact that its creation operator is
identical to its own annihilation, called the state. Its topological nature
ensures robustness against perturbations from deformations of order
parameters and nonmagnetic impurities. The states obey non-Abelian
statistics with their pairs constituting a qubit. This might offer a
promising method of the fault-tolerant quantum computation \cite{Kitaev}.
The main issue is to isolate the states from those above the minigap \cite%
{Sarma}.

The $l=1$ Majorana mode in the $p$- wave vortex states has been a popular
topic of study over the last several years \cite%
{Sigrist,Galitsky,Machida,Radzihovsky}. While the minigap in the $s$ and $d$%
-wave superconductors was detected by STM \cite{Hess}, in $p$-wave it has
not been observed. The major reason for that is the \textit{small} \textit{%
value of the} \textit{minigap} $\omega \ $in the core spectrum (just $mK$
for $Sr_{2}RuO_{4}$). To address this problem a current trend was to propose
increasingly sophisticated combinations of materials and geometries. For
example one of the proposals \cite{Sau} to expose the Majorana state is to
induce the $s$-wave superconductivity by the proximity effect on the surface
of a topological insulator. The minigaps of the resultant non-Abelian states
can be orders of magnitude larger than in a bulk chiral $p$-wave
superconductor.

In this note we propose to solve the Majorana minigap problem within the
original system, a $p_{x}+ip_{y}$ bulk superconductor in magnetic field by
pinning the vortices on artificially fabricated dielectric inclusions of the
radius comparable to the coherence length $\xi $ of the superconductor \cite%
{Nguyen}. It was shown theoretically \cite{Melnikov09} that in the $s$-wave
superconductors pinning by an inclusion of radius of just $R=0.2-0.5\xi $
changes dramatically the subgap excitation spectrum: the minigap $\omega
\sim $ $\Delta ^{2}/E_{F}$, becomes of the order of $\Delta $. In the
present note we present the spectrum and wave functions of the core
excitations in the chiral $p$-wave superconductor. The charged states for $%
R=0.1-0.4\xi $ are significantly pushed up towards $\Delta $, so that they
therefore interfere less with Majorana state. To expose the modified charged
spectrum we propose to measure heat transport along the vortex axis, see
Fig.1b. The temperature gradient between the top side and the bottom side
drives heat along the vortex axis\cite{Kopnin03}. The exponential
temperature dependence of thermal conductivity is very sensitive to the
minigap. The temperature when the material undergoes isolator - thermal
conductor crossover rises from $\Delta /100$ for unpinned vortices to $%
\Delta /10$ for inclusion radius $R=0.4\xi $.

We start with the Bogoliubov - de Gennes (BdG) equations for the $%
p_{x}+ip_{y}$ superconductor in the presence of a single pinned vortex. The
vector potential $\mathbf{A}$ in polar coordinates, $r,\varphi $, has only
an azimuthal component $A_{\varphi }\left( r\right) $ and in the London
gauge consists of the singular part $A_{\varphi }^{s}=hc/2er$ and the
regular part of the vector potential that can be neglected for\ a type II
superconductor \cite{Sigrist}. In the operator matrix form for a\ two
component amplitude the BdG equations read:

\begin{equation}
\left( 
\begin{array}{cc}
\hat{H}_{0} & L \\ 
L^{+} & -\hat{H}_{0}^{\ast }%
\end{array}%
\right) \left( 
\begin{array}{c}
u \\ 
v%
\end{array}%
\right) =E\left( 
\begin{array}{c}
u \\ 
v%
\end{array}%
\right) ,  \label{BdGeqs}
\end{equation}%
where for anisotropic dispersion%
\begin{eqnarray}
H_{0} &=&-\frac{\hbar ^{2}}{2m_{\bot }}\nabla _{\bot }^{2}-\frac{\hbar ^{2}}{%
2m_{z}}\nabla _{z}^{2}-E_{F};  \label{BdGcomp} \\
L &=&-\frac{\Delta }{k_{F}}\left\{ s\left( r\right) e^{i\varphi }\left( i%
\mathbf{\nabla }_{x}-\mathbf{\nabla }_{y}\right) \mathbf{+}\frac{1}{2}\left[
\left( i\mathbf{\nabla }_{x}-\mathbf{\nabla }_{y}\right) s\left( r\right)
e^{i\varphi }\right] \right\} ,  \notag
\end{eqnarray}%
with $\Delta $ being the "bulk gap" of order $T_{c}$. The equations, possess
the electron-hole symmetry. The Ansatz

\begin{equation}
\begin{array}{c}
u=\frac{1+i}{\sqrt{2}}f\left( r\right) e^{il\varphi }e^{ik_{z}z} \\ 
v=\frac{1-i}{\sqrt{2}}g\left( r\right) e^{i\left( l-2\right) \varphi
}e^{ik_{z}z}%
\end{array}
\label{ansatz}
\end{equation}%
converts them (for any $l$ there are radial excitation levels denoted by $n$%
) into a dimensionless form, 
\begin{eqnarray}
-\gamma \left( \frac{\partial ^{2}}{\partial r^{2}}+\frac{1}{r}\frac{%
\partial }{\partial r}-\frac{l^{2}}{r^{2}}+\frac{1}{4\gamma ^{2}}\right)
f-2\gamma \left[ s\left( r\right) \left( \frac{\partial }{\partial r}-\frac{%
l-2}{r}\right) \mathbf{+}\frac{1}{2}\left( s^{\prime }\left( r\right) -\frac{%
s\left( r\right) }{\overline{r}}\right) \right] g &=&\varepsilon _{lk_{z}n}f;
\label{BdGeq} \\
\gamma \left( \frac{\partial ^{2}}{\partial r^{2}}+\frac{1}{r}\frac{\partial 
}{\partial r}-\frac{\left( l-2\right) ^{2}}{r^{2}}+\frac{1}{4\gamma ^{2}}%
\right) g+2\gamma \left[ s\left( r\right) \left( \frac{\partial }{\partial r}%
+\frac{l}{r}\right) \mathbf{+}\frac{1}{2}\left( s^{\prime }\left( r\right) -%
\frac{s\left( r\right) }{\overline{r}}\right) \right] f &=&\varepsilon
_{lk_{z}n}g\text{,}  \notag
\end{eqnarray}
with dimensionless energy $\varepsilon _{lk_{z}n}=E_{lk_{z}n}/\Delta $. Here
distances are in units of $\xi $. We chose the order parameter as a
discontinuous function vanishing inside the core $r<R$ and $s=\tanh (r)$ for 
$r>R$. (The question of justification of using this form often used instead
of the fully self consistent approach was extensively studied in literature
on both s-wave and non-conventional pairing \cite{Gygi}, \cite{Sigrist}). In
the clean limit BCS (applicable to $SrRu_{2}O_{4}$) $\xi =\hbar k_{\bot
}/m_{\bot }\Delta $, where 
\begin{equation}
k_{\bot }^{2}/2m_{\bot }=E_{F}/\hbar ^{2}-k_{z}^{2}/2m_{z}\text{,}
\label{dispersion}
\end{equation}%
and for given $k_{z}$ there is just one dimensionless parameter 
\begin{equation}
\ \gamma =1/2k_{\perp }\xi =m_{\bot }\Delta /2\hbar ^{2}k_{\bot }^{2}\text{.}
\label{gamma}
\end{equation}

The Ansatz, Eq.(\ref{ansatz}) was chosen in such a way that the equations
become real. In the presence of a hole of radius $R$ we assume that the
order parameter profile is still accurate for $r>R$. In a microscopic theory
of the superconductor-insulator interface, (see \cite{DeGennes}), the order
parameter rises abruptly from zero in dielectric, where amplitudes of normal
excitations $f=g=0$, to a finite value inside the superconductor within an
atomic distance $a$ from the interface, namely with a slope $\propto 1/a$.
This means that the boundary condition on the amplitudes is consistent with
zero order parameter at the boundary point $r=R-a$ in the self consistency
equation. The sample will be cylindric with radius $L$, see Fig.1a.

Qualitatively the spectrum of a single vortex in a hollow disk with internal
and external radii $R$ and $L$ consists of two Majorana states, several
pairs of Andreev bound subgap states both quasiparticles and holes (related
to each other by the electron - hole symmetry) and a continuum of states
above a threshold at $T_{c}$. The spectrum was calculated numerically by
using NAG Fortran Library Routine Document F02EBF. It computes all the
eigenvalues, and optionally all the eigenvectors, of a real general matrix,
for various values of the inclusion radius $R=0\sim 0.9\xi $, external
radius $L\sim 10-20$ and the universal parameter $\gamma \left( k_{z}\right)
=1/2k_{\bot }\xi \sim 10^{-3}-3\cdot 10^{-2}$. For sufficiently small $l$,
in addition to the continuum of states above the superconducting gap with $%
n\geq 1$, there are bound Andreev states that correspond to the lowest $%
n\equiv 0$. For $l=1$ and various inclusion radii $R$ there is the Majorana
state, for which $f\left( r\right) =g\left( r\right) $ near the "internal"
surface" extending to the distance $\xi $ into the superconductor, see
Fig.2a where the radial density, 
\begin{equation}
\rho \left( r\right) =2\pi r\left( \left\vert f\left( r\right) \right\vert
^{2}+\left\vert g\left( r\right) \right\vert ^{2}\right) ,  \label{rho}
\end{equation}%
is given for sufficiently large external radius of cylinder, $L=20$ and $%
\gamma =0.03$. As was noticed in ref. \cite{Galitsky}, the function
oscillates with period $1/k_{\bot }\xi $. There is also the second Majorana
mode on the "external" surface for which $f\left( r\right) =-g\left(
r\right) $, given in Fig.2b. The energy of the both states is exponentially
small $\varepsilon _{1k_{z}}\propto e^{-L}$ due to tunneling between them%
\cite{Mizushima}. The surface state wave function practically does not
depend on the inclusion radius for all values of $R$ considered, although it
does depend on $L$. When one considers a small cylinder of the width of
several coherence length like the one shown in Fig.1a, the two Majorana
modes start to overlap. All the other $l=1$ excitations, $n\geq 1$ , are
above the threshold and will not be considered and the index omitted.

At angular momenta $l\neq 1$ topology does not protect the energy (more
precisely its absolute value). The wave function of bound states, is pushed
out of its position near the vortex core when the inclusion is absent. The
state that was localized low energy at $R=0$, becomes delocalized and
approaches the threshold at $R=0.2\xi $ and eventually at $R=0.4\xi $ merges
into the threshold. The energies in units of $\Delta $ for wide range of
angular momenta $l$ for $R=0$ (the blue points), $0.1\xi $ (green), $0.2\xi $
(red), $0.4\xi $ (brown) are given in Fig. 3. The $R=0$ line is well
approximated for $\left\vert l\right\vert <10$ by the semiclassical linear
formula \cite{Volovik99}, while for higher momenta it approaches the
threshold along a universal curve that is independent of the inclusion
radius. When the hole is present, the dependence on $l$ is no longer
monotonic. It first rises to a maximum at $l=-1$ and subsequently has a
local minimum with energy about that of the $l=0$ state. Consequently this
becomes a new minigap that becomes of order of the bulk gap already at
relatively small inclusion radius. Beyond this minimum the curve approaches
the universal $R=0$ spectrum, so that wave functions are located far from
the inclusion. Beyond radius $R=0.4\xi $ the effect of the minigap
enhancement is saturated. Only the value of $\gamma =0.03$ is shown in Fig.3
although dependence on it is very weak for all $\gamma <<1$. The enhanced
minigap allows to utilize the Majorana states.

Thermal conductivity is an effective tool to demonstrate the minigap due to
activated behavior of the electron contribution. To calculate the
quasiparticle contribution to thermal conductivity along the vortex cores
when the upper side of the vortex line is held at temperature $T_{1}$ and
the lower side at temperature $T_{2}$, see Fig.1b, we use a general
ballistic (width of the film $L_{z}$ smaller than mean free path)
Kopnin-Landauer formula\cite{Kopnin03}. The heat current at temperature
lower than the threshold to continuum of states is carried by the bound core
states (except the Majorana). For a single vortex it consists of the
contribution of quasiparticles and holes, $I=2\dsum\nolimits_{l<1}I_{l}%
\left( T_{2}\right) -I_{l}\left( T_{1}\right) $,

\begin{equation}
\text{\ }I_{l}\left( T\right) =\int_{0}^{k_{z}^{\max }}\frac{dk_{z}}{2\pi
\hbar }\left\vert \frac{dE_{lk_{z}}}{dk_{z}}\right\vert \frac{E_{lk_{z}}}{%
1+\exp \left( E_{lk_{z}}/T\right) },  \label{I}
\end{equation}%
where the energy depends on transferred momentum along the field $k_{z}$ via 
$\gamma $, see Eq.(\ref{gamma}). The maximal value of $k_{z}\ $is $%
k_{z}^{\max }=\sqrt{2m_{z}E_{F}}$. Since $E_{lk_{z}}$ is monotonic the
integral can be transformed into 
\begin{equation}
\text{ \ \ }I_{l}\left( T\right) =\int_{E_{l}^{0}}^{E_{F}}\frac{dE}{2\pi
\hbar }\frac{E}{1+\exp \left( E/T\right) }\approx \frac{T^{2}}{2\pi \hbar }%
\Pi \left( \frac{E_{l0}}{T}\right) \text{.}  \label{I1}
\end{equation}%
Since the temperatures are below the threshold, $E_{F}$ was replaced in the
upper limit of the integral by $\infty $. Here lower limit of integration is
energy for $k_{z}=0$ and 
\begin{equation}
\Pi \left( x\right) =\pi ^{2}/6-x^{2}/2+x\log \left( 1+e^{x}\right)
+Li_{2}\left( -e^{x}\right) \text{,}  \label{Pi}
\end{equation}%
where $Li$ is the polylog function. For small temperature differences the
linear response can be used,

\begin{equation}
\frac{\hbar }{\Delta }\frac{dI}{dT}=\frac{t}{\pi }\dsum\nolimits_{l<1}\left[
2\Pi \left( \frac{\varepsilon _{l0}}{t}\right) +\frac{\left( \varepsilon
_{l0}/t\right) ^{2}}{1+\exp \left( \varepsilon _{l0}/t\right) }\right] \text{%
,}  \label{minigapth}
\end{equation}%
where $t=T/\Delta $. The values of energies of the core states, presented in
Fig. 3, therefore allow to calculate the thermal conductivity. In Fig.4 the
heat conductance of a single vortex line is given as function of inverse
temperature (in units of $\Delta ^{-1}$). While for unpinned vortex (the
blue line) the crossover from thermal insulator to conductor is not well
defined in the relevant temperature range $\Delta /100<T<\Delta /4$, for the
radius of the inclusion $R=0.1\xi $, $0.2\xi $, $0.4\xi $ (brown, red and
green respectively) one observes a well defined crossover temperature.

To summarize, the wave functions and the spectrum of the core states well
pinned vortices in the chiral $p$ wave superconductor in magnetic field was
studied, see Fig. 3. The pinning by a nanohole of the order of coherence $%
\xi $ is assumed. It is shown that while the neutral Majorana mode is
largely intact, all the other Andreev bound states have their energies
significantly lifted. The minigap (the energy difference with the Majorana
state) is consequently increased from its value for unpinned vortices, $%
\Delta ^{2}/E_{F}$, to a fraction of the bulk superconducting gap $\Delta $.
In superconductor $Sr_{2}RuO_{4}$ while the minigap for unpinned vortices is
just (using \cite{Maeno} $\Delta =2K,\ E_{F}=10^{3}K$) $\omega =4mK$, it
would become about $0.2K$ for a nanohole of radius of ($\xi =65nm$ \cite%
{Maeno} ) $R=0.4\xi =25nm$. This would suppress the thermodynamic and
transport effects of the charged excitations and allow to observe the
physics of the Majorana state.

We propose to demonstrate the enhanced minigap by measuring thermal
conductivity along the vortex direction. The activated behavior of the
electron contribution per vortex is given in Fig.4 and compared to the
phonon contribution that dominates at temperatures below $T_{m}=\Delta /10$.
Magnetic field $B>>H_{c1}$ creates $N=SB/\Phi _{0}$ vortices, see Fig. 1b,
over area $S$, so that heat conductivity is $\kappa =\frac{L_{z}B}{\Phi _{0}}%
\frac{dI}{dT}$, where $L_{z}$ is the sample width. For $B=0.5T$ (between $%
H_{c1}$ and $H_{c2}$ for in $Sr_{2}RuO_{4}$), with $R=25nm;$ $L_{z}=70nm$ at 
$T=0.2K$ one obtains conductivity $\kappa =0.05W/Km$. The phonon
contribution in ballistic regime ($L_{z}<<l_{ph}$ , where $l_{ph}$ is the
phonon mean path) is \cite{Landau10} $\kappa \varpropto Cv_{ph}L_{z}$.
Taking the phonon heat capacity $C$ in Debye approximation, one obtains for
phonon conductance per vortex,%
\begin{equation}
\hbar \kappa _{ph}/\Delta =t^{3}\left( \Delta /\Theta _{D}\right)
^{3}L_{z}n_{A}\Phi _{0}/B,  \label{phonon}
\end{equation}%
where $\Theta _{D},$and $n_{A}$ are the Debye temperature, density of the
atoms. This dependence for various vortex densities presented in Fig.4
(dashed lines) demonstrates that for temperature above the minigap value,
the quasiparticle heat conductance clearly dominates. In particular for
parameters $L_{z}=50nm,B=0.5T,a=0.5nm,T_{m}=0.2$( here $a$ is the
inter-atomic distance).

It should be noted that vortices in such an experiment (for the field
cooling protocol) will be trapped by the holes rather than by point defects.
Those initially not pinned by the holes can be effectively pushed out by a
small bias current. Our experimental proposal does not require an ideal
hexagonal lattice of identical holes. Periodicity does not play a role and
the array may even be random.

\textit{Acknowledgements.} B.Ya.S. and I.S. acknowledge support from the
Israel Scientific Foundation.

\newpage

\bigskip

Figure captions

\bigskip

Fig.1a.

A single vortex in type II superconductor in magnetic field pinned on an
insulator insertion of radius $R\sim \xi $ parallel to the field. The radius
of the superconducting disk is $L>>\xi $. The geometry used to calculate the
spectrum of the chiral $p$-wave core states.

Fig. 1b.

Heat flow through a superconductor in magnetic field along the vortex cores.
The temperature difference between the bottom ($T_{2}$) and the top ($T_{1}$%
) contacts leads to energy flow carried by both the neutral (Majorana) and
charged core states. Numerous vortices are pinned in inclusions and might be
arranged as a lattice.

Fig.2a.

The radial density $\rho \left( r\right) $, Eq.(\ref{rho}), of the Majorana
core mode localized on the inclusion (green area) of radius $R=0.2\xi $ as
function of the distance from the vortex center. The value of the only
parameter characterizing the system is $\gamma =0.03$. The energy of the
state exponentially small as function of the sample radius $L$. The wave
function is pushed out of the center by the inclusion, but otherwise remains
intact compared to the unpinned vortex. It extends several coherence lengths
inside the superconductor.

Fig.2b.

The radial density $\rho \left( r\right) $ as function of the distance from
the vortex center, Eq.(\ref{rho}), of the Majorana mode localized on the
surface the sample at $r=L=20\xi $. Due to tunneling to the Majorana core
state it also has a negligible energy.

Fig. 3.

Excitation energy (in units of the bulk gap $\Delta $) of the bound states
as function of angular momentum for several values of the inclusion radius $%
R $ at $\gamma =0.03$.\bigskip\ For unpinned vortices $R=0$ (the blue
points) the spectrum at small $l$ is described well by the semiclassical
Volovik formula (see \cite{Volovik99}), while for higher momenta it
gradually approaches the threshold along a universal curve that is
independent of the inclusion radius $R$. When the inclusion is present the
dependence on $l$ is no longer monotonic. It first rises to a maximum at $%
l=-1$ and subsequently has a local minimum with energy about that of the $%
l=0 $ state.

Fig.4.

Dimensionless heat conductance\ $\frac{\hbar }{\Delta }\frac{dI}{dT}$ of a
single vortex given in Eq.(\ref{minigapth}) as function of inverse
temperature in units of $\Delta ^{-1}$. The black line is for unpinned
vortex, while the corresponding crossover temperature from thermal insulator
to conductor for pinned vortices on insulating inclusions of radius $%
R=0.1\xi $ (brown), $R=0.2\xi $ (red), $R=0.4\xi $ (green) are much higher.
Dashed likes are the contributions of phonons per vortex for $B=0.5T$
(green), $0.2T$ (black), $0.1T$ (blue), length $L_{z}=50nm$, density $%
n_{A}=10^{23}cm^{-3},\hbar \omega _{D}=400K$.

\end{document}